\documentclass[11pt, a4paper]{article}
\usepackage{geometry}                
\usepackage{graphicx}
\usepackage{amssymb}
\usepackage{epstopdf}
\begin{document}

\title{Cosmological models with linearly varying deceleration parameter}
\author{\"{O}zg\"{u}r Akarsu \footnote{E-Mail: oakarsu@ku.edu.tr}\and Tekin Dereli \footnote{E-Mail: tdereli@ku.edu.tr}}

\date{}

\maketitle
\begin{center}
\vskip-1cm
\textit{Department of Physics, Ko\c{c} University, 34450 {\.I}stanbul/Turkey.}
\end{center}

\begin{abstract}
We propose a new law for the deceleration parameter that varies linearly with time and covers Berman's law where it is constant. Our law not only allows one to generalize many exact solutions that were obtained assuming constant deceleration parameter, but also gives a better fit with data (from SNIa, BAO and CMB), particularly concerning the late time behavior of the universe. According to our law only the spatially closed and flat universes are allowed; in both cases the cosmological fluid we obtain exhibits quintom like behavior and the universe ends with a big-rip. This is a result consistent with recent cosmological observations.
\begin{flushleft}
\end{flushleft}
\end{abstract}

\section{Introduction}
\label{Intro}

It has been announced in 1998 that type Ia Supernova observations indicate that the current universe is not only expanding but also accelerating. This was perhaps the most striking discovery of the modern cosmology. Today, this behavior of the universe is confirmed by various independent observational data, including the Type Ia Supernova (SNIa), the large scale structure, the cosmic microwave background (CMB) radiation, and so on. There is a consensus on the conclusion that the universe has entered a state of accelerating expansion at redshift $z\sim0.5$. In two recent studies by Cunha and Lima \cite{Cunha08} and Cunha \cite{Cunha09}, the transition redshift to the accelerating expansion of the current universe has been examined using the kinematic approach to cosmological data analysis that provides a direct evidence to the present accelerating stage of the universe which does not depend neither on the validity of general relativity, nor on the matter-energy content of the universe. Today we can examine not only when the cosmic acceleration began and the current value of the deceleration parameter, but also how the acceleration (the deceleration parameter) varies with time. In a recent paper, Li et al. \cite{Li11} has examined the current acceleration of the universe by the largest and latest SNIa sample (Union2), baryonic acoustic oscillation (BAO) and cosmic microwave background (CMB) radiation together. They conclude that once the systematic error is taken into account, two different subsamples of Union2 along with BAO and CMB all favor an increase of the present cosmic acceleration. The authors also give plots they obtained for the change of the deceleration parameter $q$ with redshift $z<2$.

Motivated from the studies outlined above, we propose here a linearly varying deceleration parameter (LVDP), which can be used in obtaining accelerating cosmological solutions. As a special case, LVDP also covers the special law of variation for Hubble parameter, which yields constant deceleration parameter (CDP) models of the universe, presented by Berman \cite{Berman83,Berman88}. LVDP is consistent with the results of Cunha, and exhibits a similar behavior to the one obtained by Li et al. \cite{Li11} when we plot the deceleration parameter as a function of redshift. After the discovery of the late time acceleration of the universe, many authors have used CDP to obtain cosmological models in the context of dark energy (DE) in general relativity and some other modified theories of gravitation such as $f(R)$ theory within the framework of spatially isotropic and anisotropic spacetimes. However, generalizing CDP assumption would allow us to construct more precise cosmological models. Indeed, as we show in this paper, a simple generalization of CDP ansatz to LVDP ansatz fits the recent observational results pretty well. In this paper, for convenience, we consider only the spatially homogenous and isotropic Robertson-Walker (RW) spacetimes and hence the perfect fluid representation for the energy-momentum tensor in the context of general relativity. The LVDP ansatz can also be used in spatially homogenous but anisotropic spacetimes and hence the imperfect fluids might also be considered in the context of general relativity, as well as generalized theories of gravity such as $f(R)$ gravity.

\section{The field equations, law for the deceleration parameter and solution}
\label{sec2}

The Einstein field equations can be written as follows,
\begin{equation}
G_{\mu\nu} \equiv R_{\mu\nu}-\frac{1}{2}Rg_{\mu\nu}=T_{\mu\nu},
\end{equation}
where $G_{\mu\nu}$ is the Einstein tensor and $T_{\mu\nu}$ is the energy-momentum tensor. To solve the Einstein field equations it is usually necessary to make some simplifying assumptions such as choosing a metric with a significant degree of symmetry. In this paper, we consider the Robertson-Walker metric with a maximally symmetric spatial section,
\begin{equation}
ds^2=-dt^2+a^2(t)\left[\frac{\mathrm{d} r^2}{1-\kappa r^2}+r^2\,(\mathrm{d} \theta^2+\sin^2 \theta\,\mathrm{d} \phi^2 )\right],
\end{equation}
where $a(t)$ is the cosmic scale factor and the spatial curvature index $\kappa=-1,\,0,\,1$ corresponds to spatially open, flat and closed universes, respectively. Considering a co-moving fluid, the RW metric admits only the perfect fluid representation for the energy-momentum tensor, which can be written as follows,
\begin{equation}
T^{\mu}_{\nu}=[\rho,p,p,p]
\end{equation}
where $\rho=\rho(t)$ is the energy density and $p=p(t)$ is the pressure of the fluid. In a comoving coordinate system, the Einstein field equations (1) with (2) and (3), read as
\begin{equation}
\label{eqn:E1}
3\frac{{\dot{a}}^2}{a^2}+3\frac{\kappa}{a^2}=\rho,
\end{equation}
\begin{equation}
\label{eqn:E2}
\frac{{\dot{a}}^2}{a^2}+2\frac{{\ddot{a}}}{a}+\frac{\kappa}{a^2}=-p.
\end{equation}
This system consists of two equations but three unknowns ($a$, $\rho$ and $p$), hence is not fully determined. One may determine the system fully by specifying a theory that determines a relation between the energy density and pressure of the fluid. Most of the perfect fluids relevant to cosmology obey an equation of state of the form
\begin{equation}
\label{eqn:EoS}
p=w\rho,
\end{equation}
where $w$ is the equation of state (EoS) parameter, not necessarily constant. The three most common examples of cosmological fluids with constant $w$ are the dust ($w=0$), radiation ($w=\frac{1}{3}$) and vacuum energy ($w=-1$) that is mathematically equivalent to cosmological constant $\Lambda$. It is well known that fluids with $w<-\frac{1}{3}$ are usually considered in the context of DE, since they give rise to accelerating expansion. Besides the fluids with constant EoS parameter with $w<-\frac{1}{3}$, particularly in the context of DE, there have been proposed various scalar field models that can be described by a time dependent $w$ that can evolve below the $-\frac{1}{3}$, e.g., quintessence ($-1\leq w \leq 1$), phantom ($w\leq -1$), quintom that can evolve across the cosmological constant boundary $w=-1$. \footnote{See \cite{Carroll03,Copeland06,Zhao,Frieman} and references therein for further reading on DE and see \cite{Cai} for an extensive review on quintom.} However, we still do not have a perfect understanding of the EoS of the DE and sometimes it is useful to think about the Einstein field equations without specifying the theory from which EoS parameter $w$ is derived \cite{Carroll03}. Hence, as an alternative, we may seek for the solutions of the Einstein field equations under an ansatz or some ans\"{a}tze concerning the kinematics consistently with the observed kinematics of the universe. Then one can derive an EoS for the possible fluid that might be driving the expansion of the universe and particularly the current acceleration of the universe in the context of DE. We may, then, check whether such a fluid is a realistic energy-momentum source or not and try to learn about the nature of the DE. Since the twice contracted Bianchi identity (${G^{\mu\nu}}_{;\nu}=0$) assures the conservation of the energy-momentum tensor (${T^{\mu\nu}}_{;\nu}=0$), we would be sure that the fluid would satisfy the energy-momentum conservation law. Furthermore, we should require positivity of the energy density of the co-moving fluid $\rho\geq 0$ to be satisfied as a condition for our fluid to be a realistic energy-momentum source.

Following a similar line of reasoning, Berman \cite{Berman83} and Berman and Gomide \cite{Berman88} proposed in the 1980's a law of variation for Hubble parameter within the context of RW spacetimes in general relativity that yields constant deceleration parameter ($q=-\frac{\ddot{a}a}{{\dot{a}}^2}=m-1$, where $a$ is the scale factor and $m\geq 0$ is a constant). In Berman's law the deceleration parameter can get values $q\geq-1$, and since $-1\leq q <0$ corresponds to the accelerating expansion, many authors have studied cosmological models using this law in the context of DE following the discovery of current acceleration of the universe. In this paper, we propose a generalized, linearly varying deceleration parameter 
\begin{equation}
\label{eqn:qlaw}
q=-\frac{\ddot{a}a}{{\dot{a}}^2}=-kt+m-1,
\end{equation}
where $k\geq 0$ and $m\geq 0$ are constants. $k=0$ reduces (\ref{eqn:qlaw}) to the law of Berman \cite{Berman83,Berman88}. Using this law one can generalize the cosmological solutions that are obtained via CDP. Furthermore, one can obtain models that fit better the cosmological data. The universe would exhibit decelerating expansion if $q>0$, an expansion with constant rate if $q=0$, accelerating power-law expansion if $-1<q<0$, exponential expansion (also known as de Sitter expansion) if $q=-1$ and super-exponential expansion if $q<-1$ (see \cite{Carroll03,Caldwell03} for super-exponential expansion). Hence, the fastest rate of expansion under the CDP ansatz is the exponential expansion. However, with our LVDP ansatz the universe inevitably evolves into the super-exponential expansion phase unless $k=0$, which is one of the possible fates of the universe according to the cosmological observations (see \cite{Carroll03,Caldwell03,Nesseris}).

Solving (\ref{eqn:qlaw}) one obtains three different form of solutions for the scale factor:
\begin{eqnarray}
\label{eqn:scalefactor}
a=a_{1}\,e^{\frac{2}{\sqrt{m^2-2 c_{1} k}}{\rm{arctanh}}\left(\frac{kt-m}{\sqrt{m^2-2 c_{1} k}}\right)}\quad\textnormal{for}\; k> 0\;\textnormal{and}\; m\geq 0,
\end{eqnarray}
\begin{equation}
a=a_{2}(mt+c_{2})^{\frac{1}{m}}\quad\textnormal{for}\; k= 0\;\textnormal{and}\; m>0,
\end{equation}
\begin{equation}
a=a_{3}e^{c_{3}t}\quad\textnormal{for}\; k=0\;\textnormal{and}\;m=0,
\end{equation}
where $a_{1}$, $a_{2}$, $a_{3}$, $c_{1}$, $c_{2}$ and $c_{3}$ are constants of integration. The last two of these solutions are for constant $q$, and hence correspond to the solutions under CDP ansatz. We will not dwell on these but only on the first one, which is new. For convenience, in the following we consider the solution for $k>0$ and $m>0$ and omit the integration constant $c_{1}$ by setting $c_{1}=0$. By doing this, we also set the initial time of the universe to $t_{\rm{i}}=0$. The reason for considering the solution only for $k>0$ and $m>0$ is not only for simplicity but also for compatibility with the observed universe. $k>0$ means we are dealing with increasing acceleration ($\dot{q}=-k<0$). Because $t_{\rm{i}}=0$ and $k>0$, the only way to shift the deceleration parameter to values higher than $-1$ is to set $m>0$. We would like to note that here, before proceeding, if one would like the universe to commence with a decelerating expansion, one may tighten our ansatz by choosing $m>1$. Under the above considerations, (\ref{eqn:scalefactor}) is further reduced to 
\begin{eqnarray}
\label{eqn:sf}
a=a_{1}\,e^{\frac{2}{m}{\rm{arctanh}}\left(\frac{k}{m}t-1\right)}.
\end{eqnarray}
Now, using this, the Hubble parameter of the universe is obtained as follows:
\begin{equation}
\label{eqn:Hubble}
H=\frac{\dot{a}}{a}=-\frac{2}{t(kt-2m)}.
\end{equation}
Substituting the scale factor (\ref{eqn:sf}) in (\ref{eqn:E1}) and (\ref{eqn:E2}) we get the energy density 
\begin{equation}
\label{eqn:rho}
\rho=\frac{12}{t^2(kt-2m)^2}+3\frac{\kappa}{{a_{1}}^2}\,e^{-\frac{4}{m}{\rm{arctanh}}\left(\frac{k}{m}t-1\right)}
\end{equation}
and the pressure 
\begin{equation}
\label{eqn:p}
p=-8\frac{kt-m+\frac{3}{2}}{t^2(kt-2m)^2}-\frac{\kappa}{{a_{1}}^2}\,e^{-\frac{4}{m}{\rm{arctanh}}\left(\frac{k}{m}t-1\right)}.
\end{equation}
Finally, using (\ref{eqn:rho}) and (\ref{eqn:p}) in (\ref{eqn:EoS}) the EoS parameter of the fluid is obtained as follows:
\begin{equation}
\label{eqn:EoSm}
w=\frac{-8 {a_{1}}^2(kt-m+\frac{3}{2})\,e^{\frac{4}{m}{\rm{arctanh}}\left(\frac{k}{m}t-1\right)}-\kappa t^2(kt-2m)^2}{12{a_{1}}^2\,e^{\frac{4}{m}{\rm{arctanh}}\left(\frac{k}{m}t-1\right)}+3\kappa t^2(kt-2m)^2}.
\end{equation}

We also solve for the deceleration parameter $q$ as a function of the redshift $z=-1+\frac{a_{z=0}}{a}$, where $a_{z=0}$ is the present value of the scale factor:
\begin{eqnarray}
\label{eqn:qz}
q(z) = 2m-1- m \tanh\left[\frac{m}{2}\ln(z+1)-{\rm{arctanh}}\left(\frac{1+q_{z=0}}{m}-2\right)\right].
\end{eqnarray}
We know that $q_{z}=q_{z=0}=-1+m$ for CDP (i.e., the $k=0$ case in our model). Indeed, one may check that the choice $k=0$ reduces (\ref{eqn:qz}) to $q(z)=-1+m$. Note that the above relation is independent of the spatial curvature.

\section{Physical behavior of the model and observational validation}
\label{Physical analysis}
In our model the universe has finite lifetime. It starts with a big bang at $t_{\rm{i}}=0$ and ends at $t_{\rm{end}}=\frac{2m}{k}$. Both of the energy density of the fluid and the scale factor diverge in finite time as $t\rightarrow t_{\rm{end}}$. This is the big rip behavior first suggested by Caldwell et al. \cite{Caldwell03}. The universe begins with $q_{\rm{i}}=m-1$, enters into the accelerating phase ($q<0$) at $t_{\rm{t}}=\frac{m-1}{k}$ (assuming $m>1$), enters into super-exponential expansion phase ($q<-1$) at $t_{\rm{se}}=\frac{m}{k}$ and ends with $q_{\rm{end}}=-m-1$.

One may observe that positivity condition on the energy density of the fluid is valid for the spatially closed and flat universes but violated for the spatially open universe (the reason being that $\kappa=-1<0$). Hence, under the ansatz of LVDP (\ref{eqn:qlaw}), spatially closed and flat universes are possible while spatially open universe is impossible. It may be worthwhile to note that Berman \cite{Berman83} also reached the same conclusion.

The EoS parameter of the fluid exhibits different behaviors for spatially closed, flat and open models. However, in all of the three models $w=-\frac{1}{3}$ at $t=t_{\rm{t}}$ (as would be expected) and $w=-\frac{2}{3}m-1$ at $t=t_{\rm{end}}$.

Within the context of general relativity, the current cosmological data from SNIa (Supernova Legacy Survey \cite{Riess}, Gold sample of Hubble Space Telescope \cite{Astier}), CMB (WMAP, BOOMERANG) \cite{Komatsu} and large scale structure (SDSS) \cite{Eisenstein} do not rule out DE with a dynamical equation of state parameter that can evolve into the phantom region ($w<-1$). \footnote{See \cite{Carroll03,Copeland06,Zhao,Nesseris} for theoretical and observational status of crossing the phantom divide line. } This is consistent with our model. Additionally, because the observational data allows DE to pass into the phantom region, it means that the data also do not rule out the possibility of a big rip in future universe, which is also consistent with our model.

From model dependent or independent analyses of the cosmological observations in the literature \cite{Cunha08,Cunha09,Li11,Frieman,Melchiorri,Ishida,Stefania,Lima}, the transition redshift of the accelerating expansion is given by $0.3<z_{\rm{t}}<0.8$. In particular, the kinematic approach to cosmological data analysis provides a direct evidence to the present accelerating stage of the universe, which does not depend on the validity of general relativity, as well as on the matter-energy content of the universe \cite{Cunha08}. By assuming a spatially flat universe Cunha \cite{Cunha08} finds that $q_{z=0}=-0.73$ and $z_{\rm{t}}=0.49$. Now, to demonstrate how LVDP matches the observed kinematics of the universe and makes additional predictions, we first plot the cosmological parameters by choosing $m=1.6$ and $k=0.097$.

\begin{figure}[ht]
\begin{minipage}[b]{1\linewidth}
\centering
\includegraphics[width=0.6\textwidth]{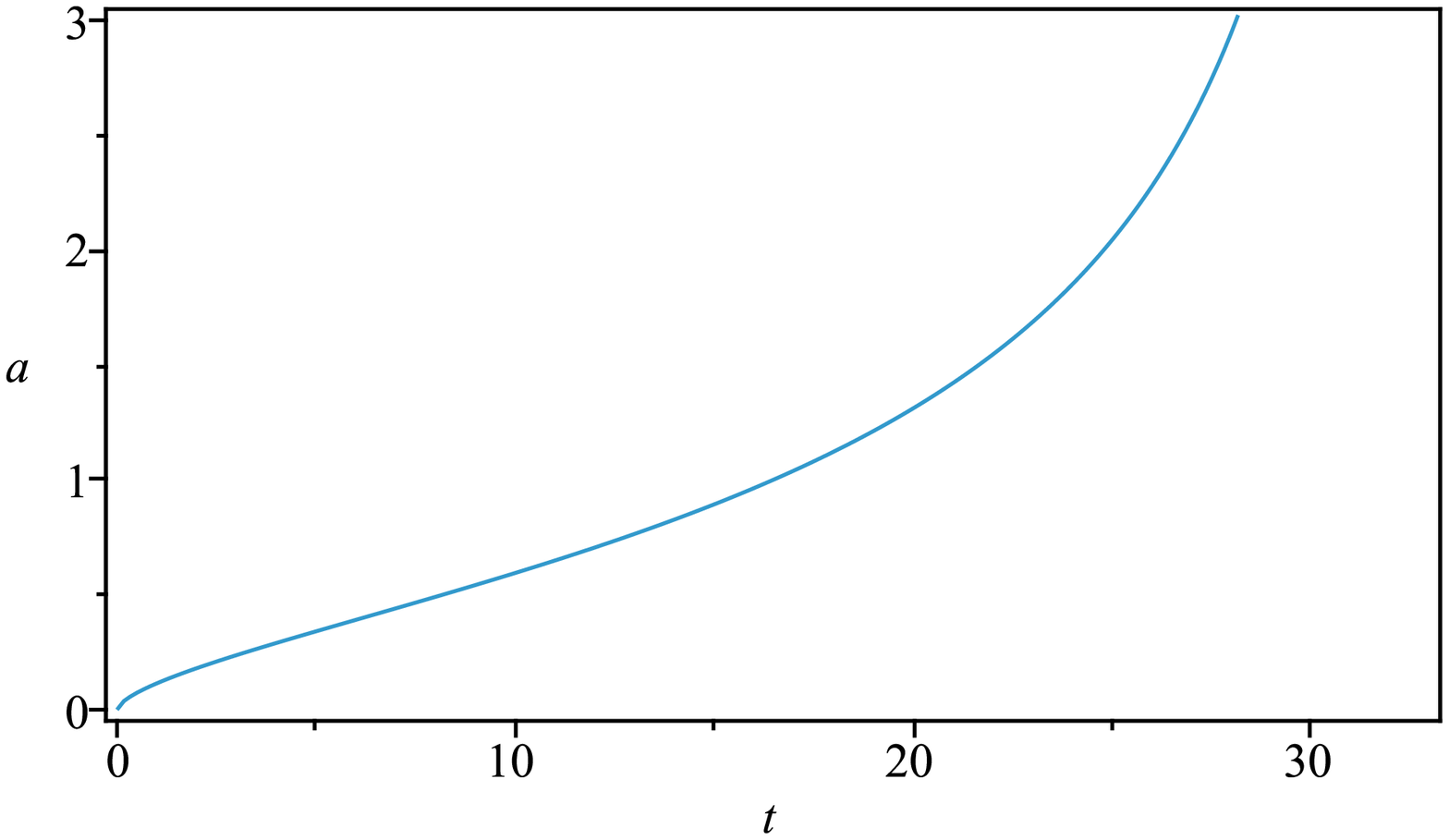}
\caption{Scale factor $a$ versus cosmic time $t$. The curve is plotted by choosing $m=1.6$ and $k=0.097$. $a$ diverges as $t\rightarrow \sim 33$.}
\label{fig:sf}
\end{minipage}
\hspace{0.01\linewidth}

\begin{minipage}[b]{1\linewidth}
\centering
\includegraphics[width=0.6\textwidth]{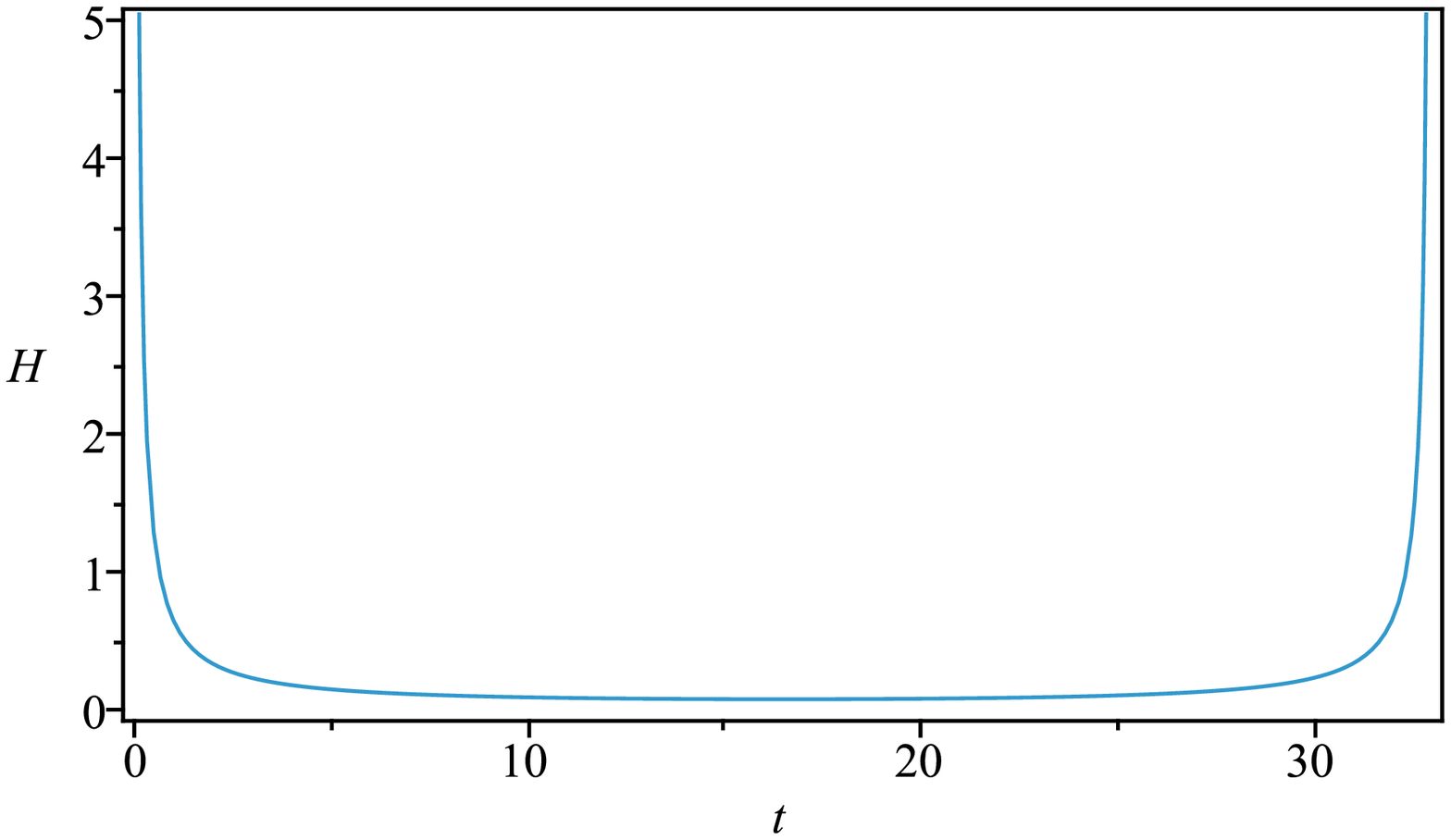}
\caption{Hubble parameter $H$ versus cosmic time $t$. The curve is plotted by choosing $m=1.6$ and $k=0.097$.}
\label{fig:h}
\end{minipage}
\hspace{0.01\linewidth}

\begin{minipage}[b]{1\linewidth}
\centering
\includegraphics[width=0.6\textwidth]{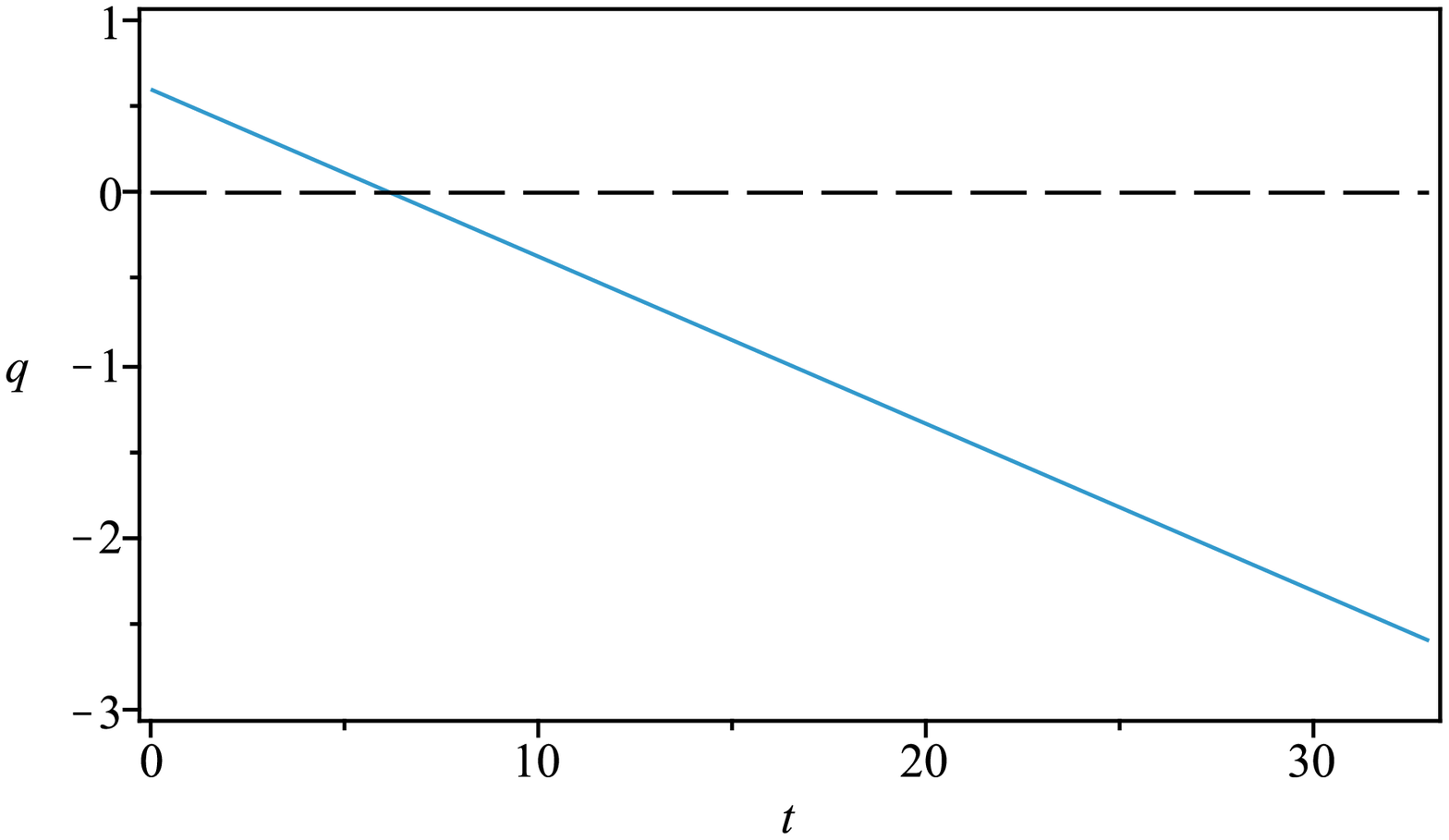}
\caption{Deceleration parameter $q$ versus cosmic time $t$ by choosing $m=1.6$ and $k=0.097$. The values of $m$ and $k$ are choosen by considering the results of Cunha \cite{Cunha09} that obtained by kinematic analysis.}
\label{fig:q}
\end{minipage}
\hspace{0.01\linewidth}
\end{figure}

In Fig. \ref{fig:sf} we plot the scale factor $a$ versus cosmic time $t$. The universe starts with big bang at $t=0$ and ends at $t_{\rm{end}}\cong 33$, which is close to the given lifetime for the universe by Caldwell \cite{Caldwell03} $t_{\rm{rip}}=35$. In Fig. \ref{fig:h} we plot the Hubble parameter $H$ versus cosmic time $t$. The Hubble parameter diverges at the beginning and end of the universe. In Fig. \ref{fig:q} we plot the deceleration parameter $q$ versus cosmic time $t$. The deceleration parameter is initially $q_{\rm{i}}=0.6$ and reaches $q_{\rm{end}}=-2.6$ at the end of the universe. The universe enters into the accelerating expansion phase at $t\cong 6.2$ and the present (i.e. at $t=13.7$) value of the deceleration parameter is $q\cong -0.73$. Both values are consistent with the observational results.

\begin{figure}[ht]
\begin{minipage}[b]{1\linewidth}
\centering
\includegraphics[width=0.55\textwidth]{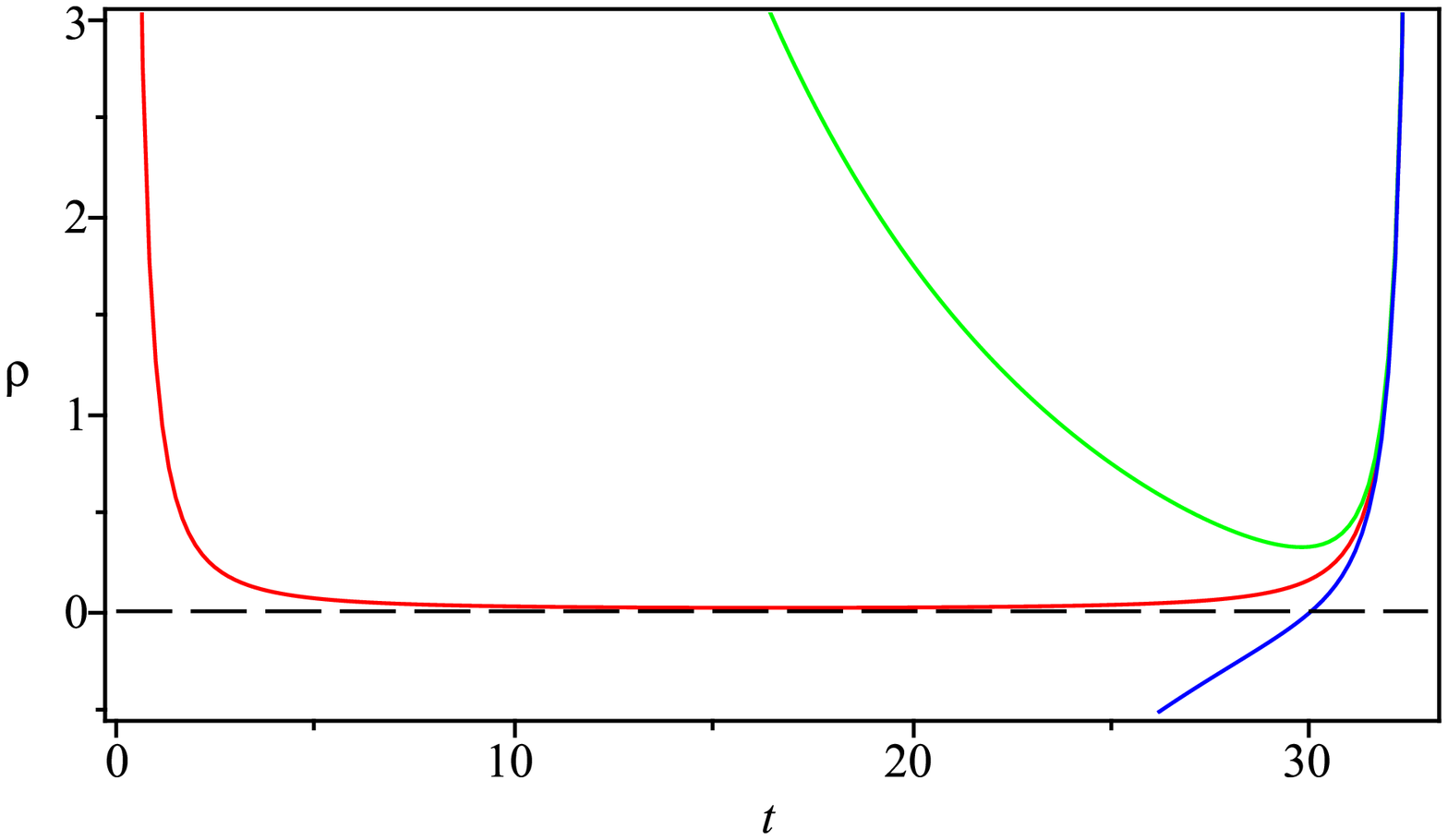}
\caption{The energy density of the fluid $\rho$ versus cosmic time $t$ for the spatially closed (green), flat (red) and open (blue) universes by choosing $m=1.6$ and $k=0.097$.}
\label{fig:rho}
\end{minipage}
\hspace{0.01\linewidth}

\begin{minipage}[b]{1\linewidth}
\centering
\includegraphics[width=0.55\textwidth]{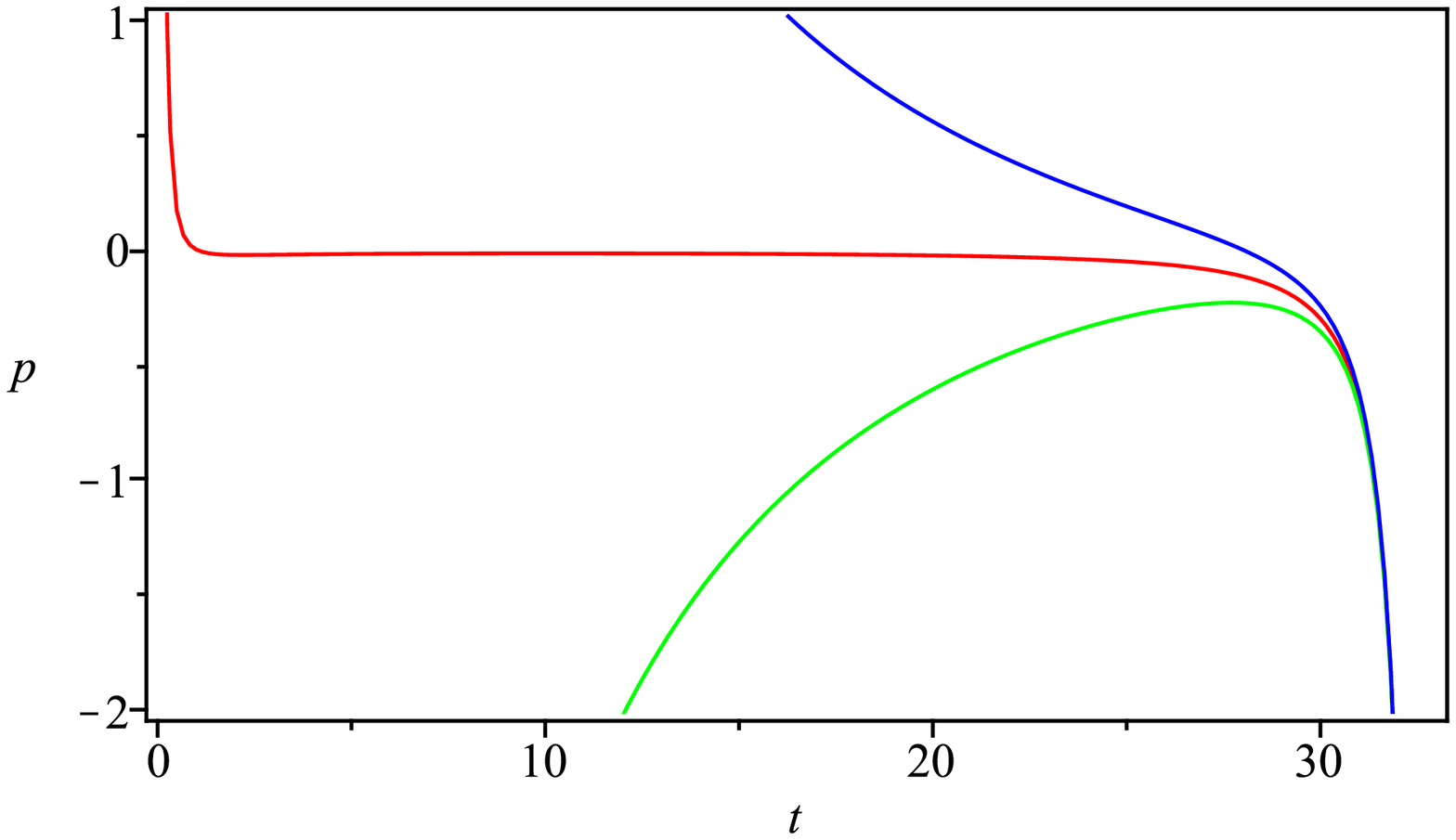}
\caption{The pressure of the fluid $p$ versus cosmic time $t$ for the spatially closed (green), flat (red) and open (blue) universes by choosing $m=1.6$ and $k=0.097$.}
\label{fig:p}
\end{minipage}
\hspace{0.01\linewidth}

\begin{minipage}[b]{1\linewidth}
\centering
\includegraphics[width=0.55\textwidth]{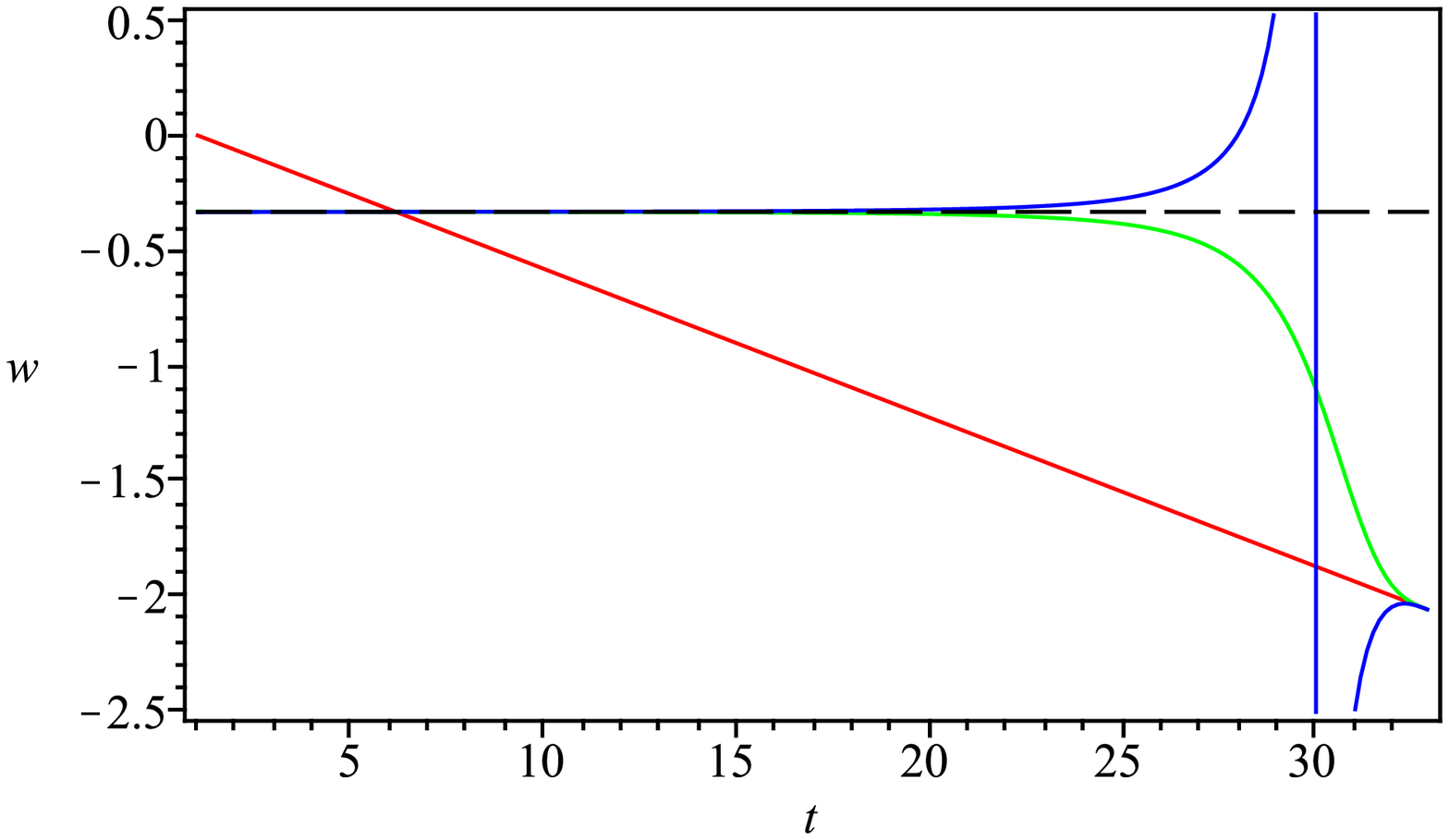}
\caption{The equation of state parameter $w$ versus cosmic time $t$ for the spatially closed (green), flat (red) and open (blue) universes by choosing $m=1.6$ and $k=0.097$. Dashed line is $w=-\frac{1}{3}$. The vertical line in the plot for spatially open universe is because the energy density of the fluid changes sign at that time.}
\label{fig:w}
\end{minipage}
\hspace{0.01\linewidth}
\end{figure}

One may observe that the spatial curvature parameter $\kappa$ is not included in the kinematics but in the dynamics of the fluid. This is what we expect, because we constrain the kinematics from the beginning but not the dynamics of the fluid. Hence, we present plots of the energy density, pressure and EoS parameter of the fluid with different colors for the spatially closed (green), flat (red) and open (blue) universes. In Fig. \ref{fig:rho} we plot the energy density of the fluid $\rho$ versus cosmic time $t$. One may observe that the spatially open model is not possible since the positivity condition of the energy density is violated but the spatially closed and flat cases are possible since the positivity of the energy density is satisfied in these models. Both for the spatially closed and flat models the energy density of the fluid diverges at the beginning and end of the universe. However, they behave differently and the energy density of the fluid in the spatially closed model is always higher than the one in the flat case. Because both of the scale factor and the energy density of the fluid diverge at $t_{\rm{end}}$, we say that the universe ends with a big rip. In Fig. \ref{fig:p} we plot the pressure of the fluid $p$ versus cosmic time $t$. The pressure also diverges at the beginning and end of the universe but exhibits different behaviours for spatially closed, flat and open universes. In Fig. \ref{fig:w} we plot the EoS parameter of the fluid $w$ versus cosmic time $t$. It exhibits different behaviors for spatially closed, flat and open universes, but converges to the same value at the very late times of the universe, strictly speaking, $w\rightarrow \sim-2$ as $t\rightarrow \sim 33$. The vertical line in the graph for the spatially open model is there since the energy density in spatially open model changes sign at that point. Hence, as already mentioned above, this solution is not possible since it violates positivity of the energy density.

\begin{figure}[ht]
\begin{minipage}[b]{1\linewidth}
\centering
\includegraphics[width=0.9\textwidth]{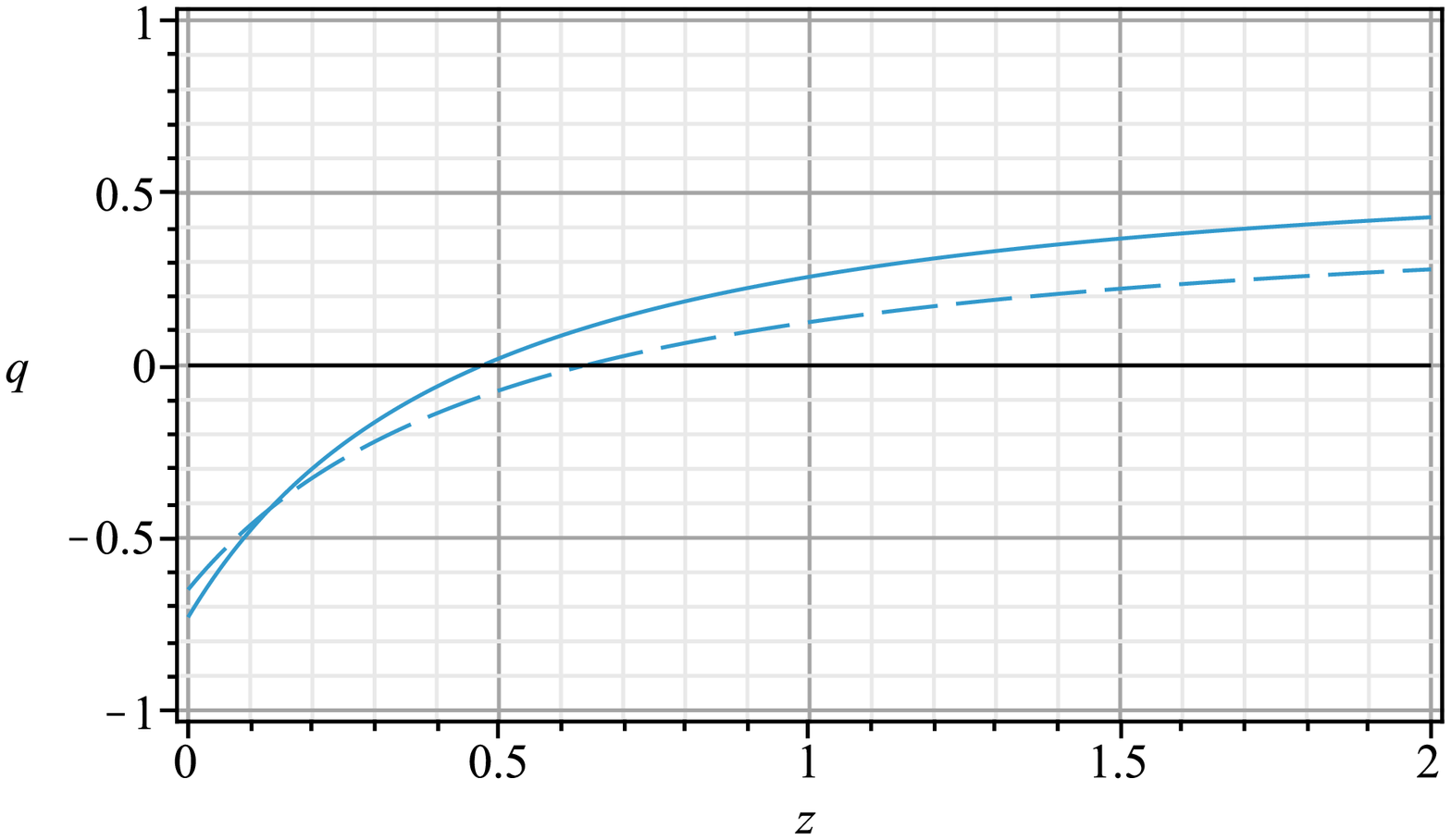}
\caption{Deceleration parameter $q$ versus redshift $z$. The solid curve is plotted by choosing $q_{{\rm{z}=0}}=-0.73$ and $m=1.6$ considering the kinematic analysis of Cunha \cite{Cunha09}. The dashed curve is plotted by choosing $q_{{\rm{z}=0}}=-0.65$ and $m=1.45$ considering the analysis of Li et al \cite{Li11}. One may check that the dashed plot fit to the graphs given by Li et al. for spatially flat universe quite well.}
\label{fig:qz}
\end{minipage}
\hspace{0.01\linewidth}
\end{figure}

We also plot the deceleration parameter $q$ versus redshift $z$ in Fig. \ref{fig:qz} with solid curve by using the same values we used in the other graphs. One may observe that the accelerating expansion begins at $z\cong 0.5$ consistent with the observational data. We particularly would like to demonstrate that by choosing $m=1.45$ and $k=0.08$ we can obtain the plot of $q(z)$ (the dashed curve in Fig. \ref{fig:qz}) which fits quite well to the plots of $q(z)$ for spatially flat universe given by Li et al. \cite{Li11} in a recent paper, where they analysed the data from SNIa, BAO and CMB simultaneously. The bottom line argument of these plots is that different data sets and analysis methods may give different values for $q_{z=0}$ and $t_{\rm{t}}$ but the general behavior of $q(z)$ follows a similar pattern. The pattern same is obtained via the LVDP ansatz that fits the observational curves quite closely.

\section{Discussion and conclusion}

We propose a special law for the deceleration parameter that is linear in time with a negative slope. The law we suggest also covers the law of Berman \cite{Berman83,Berman88} where the deceleration parameter is constant. Berman's law has been used for obtaining exact cosmological models in the literature, particularly in the context of dark energy following the discovery of the current acceleration of the universe. The law we suggest gives the opportunity to generalize many of these models with better consistency with the cosmological observations. The big-rip predicted in our models also is not ruled out by observations. The models are possible only for spatially closed and flat universe and in both models the fluid we obtained exhibits quintom like behavior.

Berman's law has also been applied to the Bianchi I spacetime assuming that the energy-momentum tensor itself is isotropic. Later on, it has been applied to various other spacetimes assuming isotropic energy-momentum tensor by many authors. Considering anisotropic spacetimes, one may also generalize the energy-momentum tensor in a way so as to yield anisotropic pressure. For instance, Akarsu and Kilinc \cite{Akarsu10} and Kumar and Singh \cite{KumarSingh} applied the law presented by Berman to Bianchi I spacetime by allowing anisotropy in the pressure of the fluid. Similarly, the law we present can also be used within the framework of spatially homogeneous but anisotropic Bianchi type spacetimes and Kantowski-Sachs spacetime in the presence of isotropic and/or anisotropic fluid. To do that, one may define the mean scale factor, most generally, as $S=(abc)^{\frac{1}{3}}$, where $a$, $b$ and $c$ are the directional scale factors, and then generalize (\ref{eqn:qlaw}) as follows,
\begin{equation}
\label{eqn:qlawgeneralized}
q=-\frac{\ddot{S}S}{\dot{S}^2}=-kt+m-1,
\end{equation}
where $k\geq 0$ and $m\geq 0$.

The law we have proposed can lead to many applications in cosmology; not only in the context of general relativity but also in generalized gravity theories such as the $f(R)$ theory of gravity.

\begin{center}
\textbf{Acknowledgments}
\end{center}
The authors appreciate the support received from the Turkish Academy of Sciences (T\"{U}BA).


\end{document}